\documentclass[11pt,a4paper]{article}
\usepackage[T1]{fontenc}
\usepackage{lmodern}
\usepackage[margin=28mm]{geometry}
\usepackage{microtype}
\usepackage{amsmath,amssymb,amsthm,mathtools}
\usepackage{booktabs}
\usepackage{tikz}
\usepackage{authblk}
\usepackage{xspace}
\usetikzlibrary{arrows.meta,positioning,bending}
\usepackage[numbers,sort&compress]{natbib}
\usepackage{xurl}
\usepackage{float}
\usepackage[colorlinks=true,linkcolor=blue!45!black,citecolor=blue!45!black,
  urlcolor=blue!45!black]{hyperref}
\newtheorem{theorem}{Theorem}
\newtheorem{lemma}[theorem]{Lemma}

\theoremstyle{definition}
\newtheorem{definition}[theorem]{Definition}
\theoremstyle{remark}
\newtheorem{remark}[theorem]{Remark}
\newcommand{\R}{\mathbb R}
\newcommand{\Z}{\mathbb Z}
\newcommand{\one}{\mathbf 1}
\newcommand{\UPAC}{\textsc{PAC-Existence}\xspace}
\newcommand{\NP}{\textsc{NP}\xspace}
\newcommand{\NPc}{\textsc{NP}-complete\xspace}
\newcommand{\PAFP}{\textsc{Path-Avoiding-Pairs}\xspace}
\newcommand{\rev}{\rightleftarrows}

\title{Detecting autocatalytic cores:\\
a simple proof of NP-completeness}
\author[ ]{Léo-Paul Bendeks}
\author[1]{Denis Kuperberg}
\affil[1]{CNRS, LIP, ENS Lyon}
\date{}

\begin{document}
\maketitle
\begin{abstract}
We give a simple proof that detecting the presence of an autocatalytic core in a reaction network is NP-complete, even when all reactions are reversible. This answers an open question of Kosc et al.~\cite{kosc2025thermodynamic}. The proof is a reduction from an undirected path problem with forbidden pairs.
The autocatalytic cores used in the reduction are simple loops with a forked reaction, corresponding to type-I autocatalytic cores in the classification of \cite{blokhuis2020universal}.
\end{abstract}

\section{Introduction}

Autocatalysis has attracted sustained interest in research on the origin of
life as a possible route from prebiotic chemistry to systems capable of
Darwinian evolution~\cite{kauffman1986autocatalytic,peng2022hierarchical,baum2023ecology,kosc2025thermodynamic}.
Following \cite{blokhuis2020universal}, we will define an autocatalytic cycle as a subnetwork in which there is a regime allowing a net production of each internal species. Reactants and products outside of the cycle are allowed, and are considered as food and waste, respectively.
As a basic form of self-amplification, autocatalytic cycles are investigated as a way to understand a possible transition from chemistry to biology.
Autocatalytic cores are minimal subnetworks with this property, so their detection is equivalent to that of autocatalytic cycles. Moreover, they can be thought of as the ``irreducible'' building blocks of autocatalysis, so identifying and listing them is a way to understand the structure of the different autocatalytic phenomena taking place in a reaction network.
The question of the computational complexity of this detection problem was only partially addressed by Kosc et al.~\cite{kosc2025thermodynamic}, leaving the general problem open. 

We will keep the terminology of Kosc et al.~\cite{kosc2025thermodynamic}, using the term \emph{potential autocatalytic core} (PAC) to refer to a minimal subnetwork that can produce each of its internal species at a strictly positive net rate. Our main result is a proof that detecting the presence of a PAC in a given reaction network is NP-complete, even when all reactions are reversible and no constraints are placed on target species or allowed food.
Here ``potential'' emphasizes the fact that there exists some regime (a vector of reaction rates) ensuring a net production of all species of the PAC, but without guarantee that this regime can actually occur in real-world conditions.

\subsection*{Related work}
Kosc et al.~\cite[Section~2.1]{kosc2025thermodynamic} proved that detecting
a PAC that contains a specified species and uses food only from a specified
set is \NP-complete. They left the unconstrained problem open. Earlier work
of Andersen et al.~\cite{andersen2012maximizing} established NP-completeness for a different notion of autocatalysis, namely the problem of maximizing the production of a specific resource. In particular, as detailed in \cite{kosc2025thermodynamic}, the reduction of Andersen et al.~\cite{andersen2012maximizing} is not about detecting a PAC, but about using a subnetwork to produce food allowing to run a known simple PAC. So autocatalysis is simply a by-product of their construction, which is really about maximizing the output of a reaction network.

A public preprint~\cite{unconstrained2026} already
claims to prove our \NP-completeness result. It is AI-generated and does not list any human author in the document, although the webpage mentions Michael Mislan as contributor. The preprint reports a Lean formalization, increasing the confidence in the result. However, the proof is less accessible than it could be: it uses elaborate gadgets, and the writing style is typical of AI, with a flow that is sometimes alien to a human way of thinking. Although AI was instrumental as well in the production of the present paper, we believe that in addition to our argument being simpler, the human contribution is valuable, both for clarity of the paper and for subsequent dissemination of the result.

\subsection*{AI disclosure}

LLMs were used in an interactive way to navigate the literature, come up with the reduction, and draft the text. The authors heavily revised the draft to ensure human readability, correctness, and clarity. We take responsibility for the content of the manuscript.

\section{Reaction networks and certificates}\label{sec:definitions}

\begin{definition}\label{def:network}
Let $E$ be a finite set of species and $R$ a finite set of reaction names. A reaction network is given by a net stoichiometric matrix
$S\in\Z^{E\times R}$, with rows indexed by species and columns by reactions. Negative entries give reactant coefficients and positive entries
give product coefficients, so reaction $r$ is written
\[
 \sum_{x:S_{xr}<0}(-S_{xr})x\ \rev\ \sum_{x:S_{xr}>0}S_{xr}x.
\]
\end{definition}

The reference orientation is left to right. A flow $v_r\in\R$ for reaction $r$ is signed:
a negative value means that the reaction is proceeding from right to left. All reactions are assumed to be reversible, so the flow can be chosen in either direction.
Given a flow vector $\vec v\in\R^R$, the net production of each species is given by the vector $S\vec v\in\R^E$. A positive entry means that the species is being produced, while a negative entry means that it is being consumed. 

We define now the autocatalytic cycles we will be interested in. Intuitively, a PAC is a minimal subnetwork where all species act as reactants and products, and in which there is a regime allowing a net production of each species.

\begin{definition}\label{def:pac}\cite{blokhuis2020universal,kosc2025thermodynamic}
A \emph{potential autocatalytic core} (PAC) is a pair of nonempty subsets
$X\subseteq E$ and $Q\subseteq R$ satisfying:
\begin{enumerate}
\item for every $r\in Q$, there are $x,y\in X$ such that
      $S_{xr}<0<S_{yr}$. This means that every selected reaction consumes and produces at least one internal species.
\item Writing $S'$ for the submatrix of $S$ restricted to rows in $X$ and columns in $Q$, there is $\vec v\in\R^Q$ such that all components of the vector $S'\vec v$ are strictly positive.
\end{enumerate}
The term "core" requires that the pair must also be minimal: no different pair of nonempty sets
$(X',Q')$ with $X'\subseteq X$ and $Q'\subseteq Q$ satisfies both
conditions.
\end{definition}

\paragraph{Example.}
Consider the reversible reactions
\[
 r_1:\ A+F\rev B,\qquad
 r_2:\ B\rev C,\qquad
 r_3:\ C\rev 2A+W.
\]
The matrix $S$ of the whole network is:
\[
 \bordermatrix{
   &r_1&r_2&r_3\cr
  A&-1&0&2\cr
  B&1&-1&0\cr
  C&0&1&-1\cr
  F&-1&0&0\cr
  W&0&0&1\cr
 }.
\]
Take $X=\{A,B,C\}$ and $Q=\{r_1,r_2,r_3\}$, leaving food $F$ and
waste $W$ external. Each reaction consumes and produces an internal
species. The flow $\vec v=(5,4,3)^{\mathsf T}$
gives
\[
 S'\vec v=
 \begin{pmatrix}
 -1&0&2\\
 1&-1&0\\
 0&1&-1
 \end{pmatrix}
 \begin{pmatrix}5\\4\\3\end{pmatrix}
 =\begin{pmatrix}1\\1\\1\end{pmatrix}>0.
\]
Moreover, it is clear that no proper subset of $X$ or $Q$ satisfies both conditions. Thus, $(X,Q)$ forms a PAC.

We will now be interested in the following decision problem:
\begin{definition} We call \UPAC the problem of deciding whether a given reaction network contains a PAC. The input is a stoichiometric matrix $S\in\Z^{E\times R}$ with coefficients given in binary encoding, and the output is \textsc{yes} if there exists a PAC $(X,Q)$, and \textsc{no} otherwise.
\end{definition}

Our main result is the following:

\begin{theorem}\label{thm:npcomplete}
\UPAC is \NPc.
\end{theorem}

\begin{remark}\label{rem:minimality}
Removing the minimality condition from Definition~\ref{def:pac} would yield an equivalent problem, since a network contains an autocatalytic set if and only if it contains a minimal one. We chose to keep the minimality condition because the previous articles \cite{blokhuis2020universal,kosc2025thermodynamic} focused on the structure and behaviours of autocatalytic cores.
\end{remark}

Membership in \NP was already mentioned in \cite{kosc2025thermodynamic}, relying on the linear algebra formulation of \cite{blokhuis2020universal}. We give here a quick proof for self-containedness. 
We rely on the following classical result:
\begin{theorem}[Khachiyan~{\cite{khachiyan1980polynomial}}]\label{thm:lp}
Given $A\in\mathbb Q^{m\times n}$ and $\vec b\in\mathbb Q^m$, one can
decide whether $A\vec x\ge\vec b$ has a solution $\vec x\in\R^n$ in time
polynomial in the total binary encoding length of $A$ and $\vec b$.
\end{theorem}

\begin{lemma}\label{lem:np}
\UPAC is in \NP.
\end{lemma}
\begin{proof}
A non-deterministic algorithm can guess nonempty sets $X\subseteq E$ and $Q\subseteq R$ and check the first item of Definition~\ref{def:pac} in polynomial time.
Strict production is equivalent, by scaling, to the existence of $\vec v\in \mathbb R^Q$ such that
$S'\vec v\ge\one$, which can be decided in polynomial time by
Theorem~\ref{thm:lp}. Minimality can be omitted by Remark \ref{rem:minimality}.
\end{proof}

The main contribution of this paper is the NP-hardness proof.

We will proceed by a reduction from a path problem in graphs with forbidden pairs, that we describe in the next section.

\section{Undirected paths with disjoint forbidden pairs}\label{sec:paths}

Our graphs here are undirected and simple, meaning that the edges have no direction, there is at most one edge between two vertices, and there are no self-loops.
The problem $\PAFP$ takes an undirected simple graph $G=(V,F)$,
source and target vertices $s\neq t\in V$, and a collection $\mathcal P$ of disjoint forbidden vertex pairs, that do not involve $s,t$. It asks for a path from $s$ to $t$ containing at most one vertex from each pair. We call such a path \emph{safe}.
See e.g. \cite{Szeider2003,Kolman2009} for several variants of the problem.

\begin{lemma}\label{lem:pafp}
$\PAFP$ is NP-complete.
\end{lemma}
\begin{proof}
This is essentially a specialization of \cite[Prop. 2]{Szeider2003} to our setting.

The problem is clearly in \NP. For \NP-hardness we reduce from 3SAT.

Let $\varphi=C_1\land\cdots\land C_m$ with $m\ge1$. Each clause $C_i$ is of the form $C_i = L_{i,1} \vee L_{i,2} \vee L_{i,3}$, where each $L_{i,j}$ is a literal, ie a variable or its negation. The graph is a chain of clauses. Clause $i$ has an
entrance $p_i$, an exit $q_i$, and three occurrence vertices
$z_{i,1},z_{i,2},z_{i,3}$, joined by the paths
\[
 p_i-z_{i,j}-q_i\qquad(j=1,2,3).
\]
Consecutive clauses are joined by the edge $q_ip_{i+1}$, and we set
$s=p_1$, $t=q_m$. Two occurrence vertices form a forbidden pair when
their literals are a variable and its negation. These pairs need not be disjoint yet.
Figure~\ref{fig:pafp-clauses} illustrates the clause chain and a safe path.

\begin{figure}[htbp]
\centering
\begin{tikzpicture}[
  occurrence/.style={circle,draw,fill=white,inner sep=2pt,minimum size=7mm},
  junction/.style={circle,draw,fill=white,inner sep=0pt,minimum size=3mm},
  chosen/.style={very thick},
  forbidden/.style={densely dashed,thick,red!60!black},
  font=\small]
\node[junction,label=below:{$s=p_1$}] (p1) at (0,0) {};
\node[junction,label=below:{$q_1$}] (q1) at (3,0) {};
\node[junction,label=below:{$p_2$}] (p2) at (4.2,0) {};
\node[junction,label=below:{$q_2$}] (q2) at (7.2,0) {};
\node[junction,label=below:{$p_3$}] (p3) at (8.4,0) {};
\node[junction,label=below:{$q_3=t$}] (q3) at (11.4,0) {};
\node[occurrence] (z11) at (1.5,1.1) {$x$};
\node[occurrence] (z12) at (1.5,0) {$y$};
\node[occurrence] (z13) at (1.5,-1.1) {$w$};
\node[occurrence] (z21) at (5.7,1.1) {$\neg x$};
\node[occurrence] (z22) at (5.7,0) {$u$};
\node[occurrence] (z23) at (5.7,-1.1) {$v$};
\node[occurrence] (z31) at (9.9,1.1) {$\neg x$};
\node[occurrence] (z32) at (9.9,0) {$y$};
\node[occurrence] (z33) at (9.9,-1.1) {$v$};
\draw (p1)--(z11)--(q1) (p1)--(z13)--(q1);
\draw[chosen] (p1)--(z12)--(q1)--(p2)--(z21)--(q2)--(p3)--(z32)--(q3);
\draw (p2)--(z22)--(q2) (p2)--(z23)--(q2);
\draw (p3)--(z31)--(q3) (p3)--(z33)--(q3);
\draw[forbidden] (z11) .. controls (2.2,2.5) and (5,2.5) ..
  node[midway,above] {$P_1$} (z21);
\draw[forbidden] (z11) .. controls (2.2,3.8) and (9.2,3.8) ..
  node[midway,above] {$P_2$} (z31);
\node at (1.5,-1.8) {$C_1$};
\node at (5.7,-1.8) {$C_2$};
\node at (9.9,-1.8) {$C_3$};
\end{tikzpicture}
\caption{The clause chain for $C_1=(x\vee y\vee w)$,
$C_2=(\neg x\vee u\vee v)$, and $C_3=(\neg x\vee y\vee v)$.
Dashed curves mark the two forbidden pairs, which share the $x$ occurrence.
Bold edges form a safe $s$--$t$ path, choosing one occurrence per clause.}
\label{fig:pafp-clauses}
\end{figure}
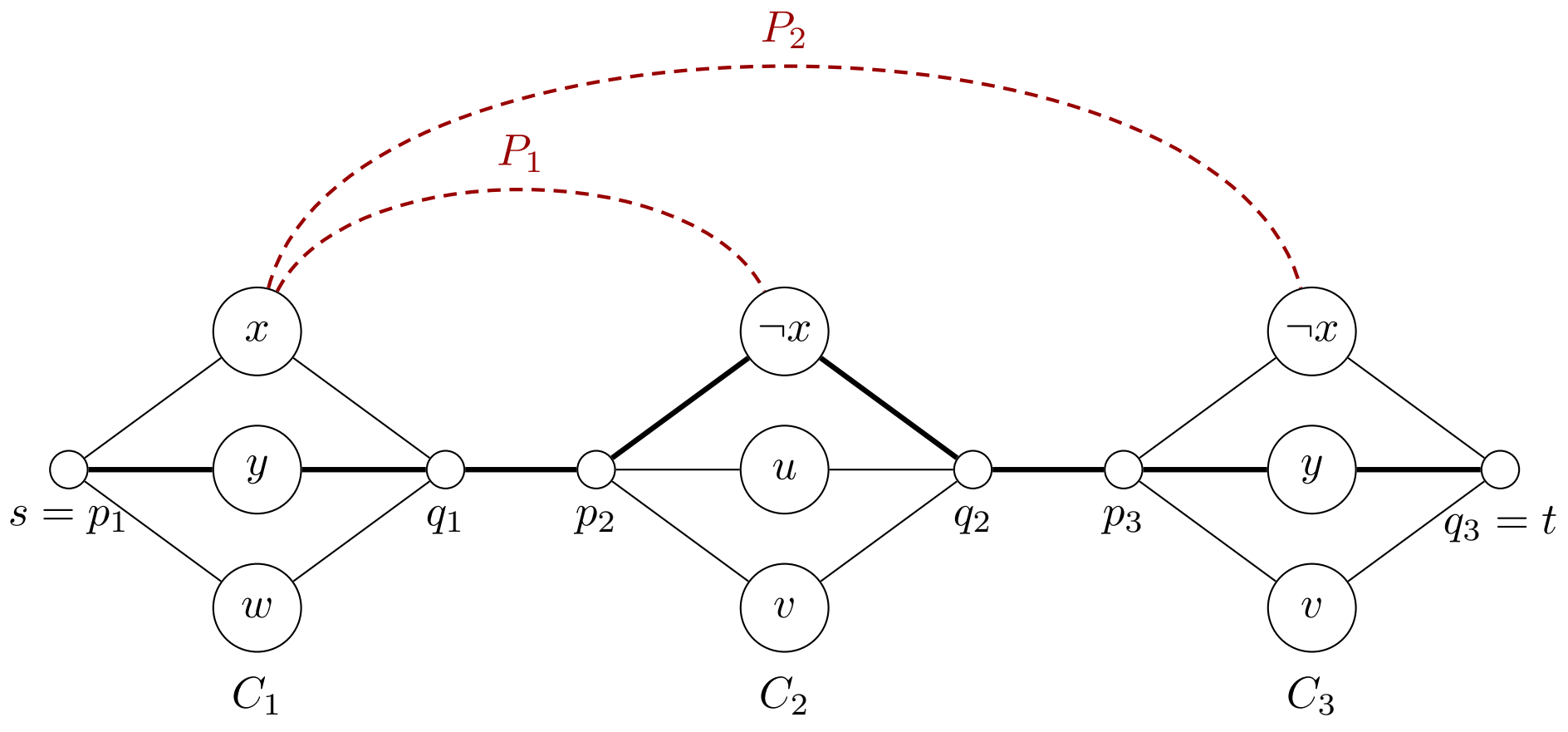

It is straightforward to observe that $\varphi$ is satisfiable if and only if there is a safe $s$--$t$ path in this graph. Indeed, a safe path must choose one occurrence per clause, and the chosen occurrences correspond to a satisfying assignment, whose consistency is guaranteed by the forbidden pairs. Conversely, a satisfying assignment allows to choose one literal occurrence per clause, yielding a safe path.

It remains to separate overlapping pairs. Let $z$ be an occurrence vertex
with neighbors $u$ and $w$, lying in the distinct forbidden pairs
$P_1,\ldots,P_k$. If $k\ge1$, replace the branch $u-z-w$ by the chain
\begin{equation}\label{eq:clones}
 u-z^{P_1}-z^{P_2}-\cdots-z^{P_k}-w,
\end{equation}
and call each $z^{P_i}$ a clone of $z$. If $k=0$, leave $z$ unchanged.
Do this for every occurrence at once, and replace each original pair
$P=\{z,z'\}$ by $\{z^P,(z')^P\}$. Every clone then belongs to exactly one
forbidden pair, and neither terminal lies in a pair.
Figure~\ref{fig:pafp-clauses2} shows the result of this transformation on the clause chain of Figure~\ref{fig:pafp-clauses}. The safe path is unchanged, and the two forbidden pairs are now disjoint.

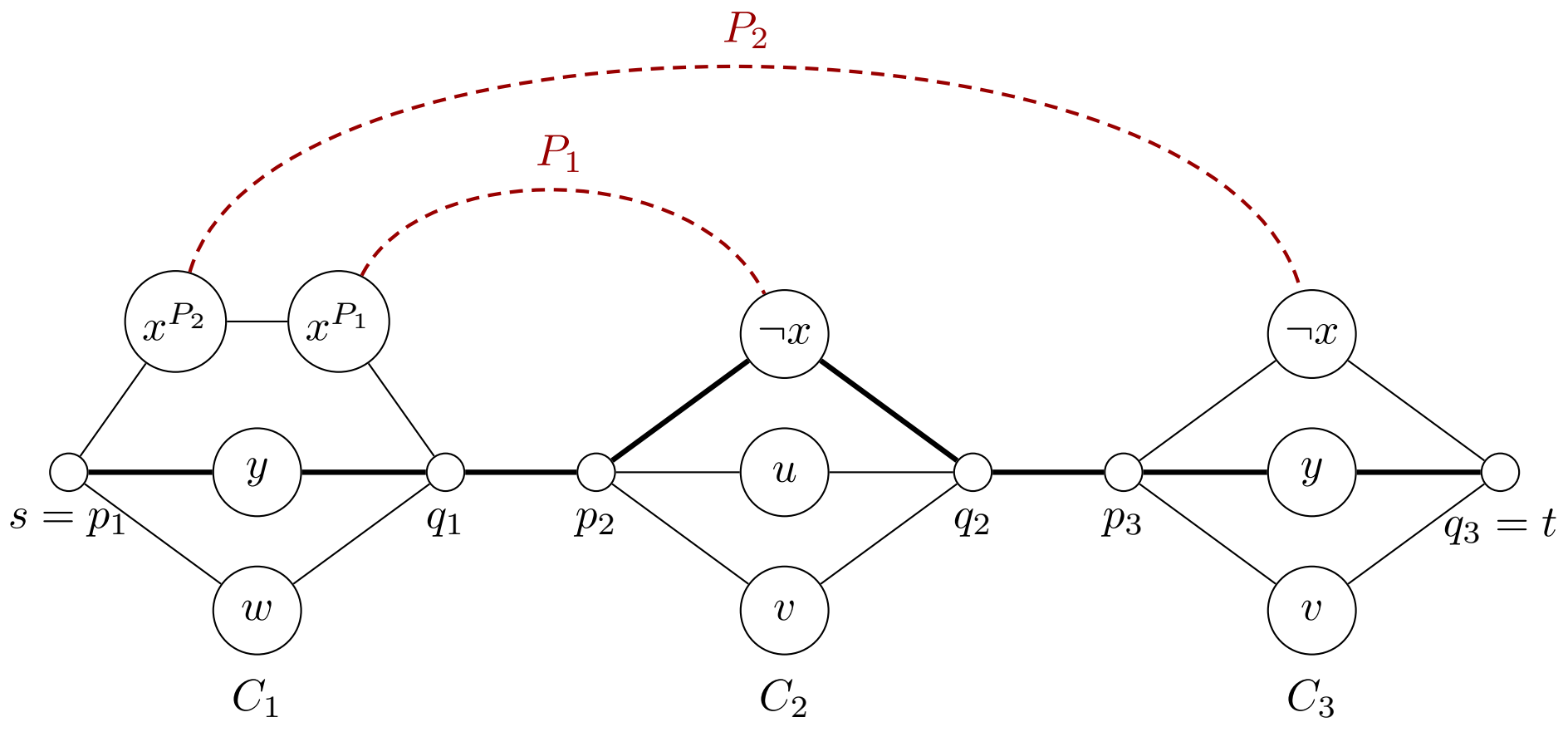
\begin{figure}[H]
\centering
\begin{tikzpicture}[
  occurrence/.style={circle,draw,fill=white,inner sep=2pt,minimum size=7mm},
  junction/.style={circle,draw,fill=white,inner sep=0pt,minimum size=3mm},
  chosen/.style={very thick},
  forbidden/.style={densely dashed,thick,red!60!black},
  font=\small]
\node[junction,label=below:{$s=p_1$}] (p1) at (0,0) {};
\node[junction,label=below:{$q_1$}] (q1) at (3,0) {};
\node[junction,label=below:{$p_2$}] (p2) at (4.2,0) {};
\node[junction,label=below:{$q_2$}] (q2) at (7.2,0) {};
\node[junction,label=below:{$p_3$}] (p3) at (8.4,0) {};
\node[junction,label=below:{$q_3=t$}] (q3) at (11.4,0) {};
\node[occurrence] (z11P2) at (.85,1.2) {$x^{P_2}$};
\node[occurrence] (z11P1) at (2.15,1.2) {$x^{P_1}$};
\node[occurrence] (z12) at (1.5,0) {$y$};
\node[occurrence] (z13) at (1.5,-1.1) {$w$};
\node[occurrence] (z21) at (5.7,1.1) {$\neg x$};
\node[occurrence] (z22) at (5.7,0) {$u$};
\node[occurrence] (z23) at (5.7,-1.1) {$v$};
\node[occurrence] (z31) at (9.9,1.1) {$\neg x$};
\node[occurrence] (z32) at (9.9,0) {$y$};
\node[occurrence] (z33) at (9.9,-1.1) {$v$};
\draw (p1)--(z11P2)--(z11P1)--(q1) (p1)--(z13)--(q1);
\draw[chosen] (p1)--(z12)--(q1)--(p2)--(z21)--(q2)--(p3)--(z32)--(q3);
\draw (p2)--(z22)--(q2) (p2)--(z23)--(q2);
\draw (p3)--(z31)--(q3) (p3)--(z33)--(q3);
\draw[forbidden] (z11P1) .. controls (2.8,2.5) and (5,2.5) ..
  node[midway,above] {$P_1$} (z21);
\draw[forbidden] (z11P2) .. controls (1.6,3.8) and (9.2,3.8) ..
  node[midway,above] {$P_2$} (z31);
\node at (1.5,-1.8) {$C_1$};
\node at (5.7,-1.8) {$C_2$};
\node at (9.9,-1.8) {$C_3$};
\end{tikzpicture}
\caption{The graph after separating overlapping forbidden pairs. The order of $x^{P_1}$ and $x^{P_2}$ was chosen purely for visual clarity.}
\label{fig:pafp-clauses2}
\end{figure}

After this transformation, we obtain a graph with disjoint forbidden pairs, and the original instance has a safe path if and only if the transformed instance does.

There are $3m$ occurrences and at most $\binom{3m}{2}$ original pairs.
There are twice as many clones as pairs, and $O(m)$  other vertices, so the instance has size $O(m^2)$ and can be built in polynomial time. This completes the \NP-hardness proof of $\PAFP$.
\end{proof}

\section{The reduction from $\PAFP$ to \UPAC}\label{sec:construction}

We will now reduce $\PAFP$ to \UPAC, in order to show that \UPAC is \NP-hard. 

Let us fix an instance $I=(G,s,t,\mathcal P)$ of $\PAFP$, with $G=(V,F)$, $s,t\in V$, and $\mathcal P$ a collection of disjoint forbidden pairs in $V$. We will build a reaction network $N_I$ such that $N_I$ contains a PAC if and only if $G$ has a safe $s$--$t$ path.

We write $\bar u$ for the vertex paired with $u$, if it exists. We define a species for each vertex $u\in V$ and a species for each edge $e\in F$: $E_I=V\uplus F.$
We will simply write $u$ and $e$ for these species, and specify whenever needed whether the notation points to the species or the corresponding vertex/edge.

For each vertex $u$ lying on an edge $e$, add the \emph{incidence reaction}
\begin{equation}\label{eq:incidence}
 r_{u,e}:\quad
 \begin{cases}
 u\rev e+\bar u,&\text{if $u$ is paired},\\
 u\rev e,&\text{if $u$ is unpaired}.
 \end{cases}
\end{equation}
Thus each edge gives two reactions. Add one more, the \emph{amplifier},
\begin{equation}\label{eq:amplifier}
 r_*:\quad t\rev 2s.
\end{equation}
The resulting reaction network is $N_I$. It has $|V|+|F|$ species and $2|F|+1$ reactions, and can be built in polynomial time from the \PAFP instance $I$.

\begin{remark}\label{rem:bounds}
Every reaction in the construction has one of the forms
\begin{equation}\label{eq:types}
 A\rev B,\qquad A\rev B+C,\qquad A\rev 2B,
\end{equation}
\end{remark}

The rest of the section is devoted to proving the following lemma, which will imply Theorem~\ref{thm:npcomplete}.
\begin{lemma}
\label{lem:reduction}
The instance $I$ has a safe $s$--$t$ path if and only if the corresponding reaction network $N_I$ contains a PAC.
\end{lemma}
We will in particular show that the PACs in $N_I$ can only be of type-I in the classification of Blokhuis et al.~\cite[Fig.~2A]{blokhuis2020universal}. More precisely they consist of paths $s\to e_1\to u_1\to e_2\to\cdots\to u_{k-1}\to e_k\to t$ of incidence reactions, closed by the amplifier $t\to 2s$.

\begin{lemma}\label{lem:forward}
If $G$ has a safe $s$--$t$ path, then $N_I$ contains a PAC.
\end{lemma}
\begin{proof}
Let $s=u_0,e_1,u_1,\ldots,e_k,u_k=t$ be a safe $s$--$t$ path in $G$. 

We show that the subnetwork of $N_I$ restricted to the selected species and reactions:
\[
\begin{array}{ccc}
  X&=&\{s,u_1,\ldots,u_{k-1},t\}\cup\{e_1,\ldots,e_k\}\\
  Q&=&\{r_{u_{i-1},e_i},r_{u_i,e_{i}}:1\leq i\leq k\}\cup\{r_*\}
\end{array}
\]
is a PAC.

Indeed, it is of the simplest possible form for a PAC: a simple cycle with a forked reaction $t\to 2s$, corresponding to type-I in the classification of Blokhuis et al.~\cite[Fig.~2A]{blokhuis2020universal}. Minimality is clear, since removing any species or reaction would disconnect the cycle and produce a subnetwork where some species is either never produced or never consumed, violating Definition \ref{def:pac}.

The forbidden pairs ensure that it is actually of this form: the $\bar{u}$ that appears in the incidence reaction $r_{u,e}$ are not selected as internal species, and are therefore allowed to appear as food or waste.

Let $n=|X|=|Q|$ be the number of selected species, equal to the number of reactions of the PAC\footnote{It is shown in \cite{blokhuis2020universal} that this is always the case that a PAC has as many species as reactions.}. Let us describe a witness vector showing that this $(X,Q)$ is indeed a PAC. We assign a flow of $n$ to the amplifier $t\to 2s$, and for the $i^\mathrm{th}$ reaction of the path $s\to t$ is assigned flow $2n-i$, as shown on Figure~\ref{fig:example}. The net production of each species from $X$ is then $1$.

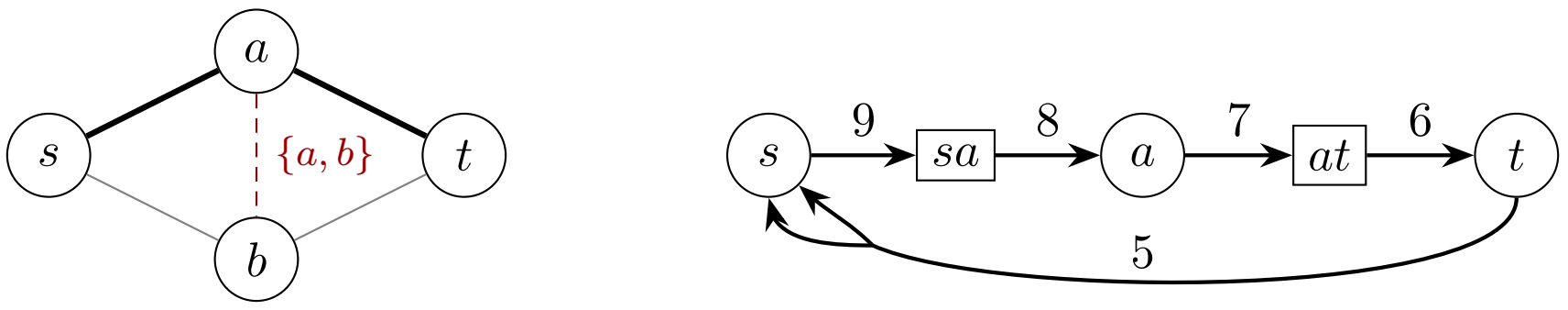
\begin{figure}[H]
\centering
\begin{tikzpicture}[
  vertex/.style={circle,draw,inner sep=2pt,minimum size=6mm},
  edgevertex/.style={rectangle,draw,inner sep=3pt},
  >=Stealth, font=\small]
\node[vertex] (gs) at (0,0) {$s$};
\node[vertex] (ga) at (1.5,.75) {$a$};
\node[vertex] (gb) at (1.5,-.75) {$b$};
\node[vertex] (gt) at (3,0) {$t$};
\draw[very thick] (gs)--(ga)--(gt);
\draw[gray] (gs)--(gb)--(gt);
\draw[densely dashed,red!60!black] (ga)--(gb)
  node[midway,right,font=\scriptsize] {$\{a,b\}$};
\begin{scope}[xshift=5.2cm]
\node[vertex] (s) at (0,0) {$s$};
\node[edgevertex] (h1) at (1.35,0) {${sa}$};
\node[vertex] (a) at (2.7,0) {$a$};
\node[edgevertex] (h2) at (4.05,0) {${at}$};
\node[vertex] (t) at (5.4,0) {$t$};
\draw[->,thick] (s)--node[above] {$9$}(h1);
\draw[->,thick] (h1)--node[above] {$8$}(a);
\draw[->,thick] (a)--node[above] {$7$}(h2);
\draw[->,thick] (h2)--node[above] {$6$}(t);
\coordinate (fork) at (.75,-.65);
\draw[thick] (t.south) .. controls (5.4,-1.05) and (1.65,-1.05) .. (fork);
\draw[->,thick] (fork) to[out=180,in=-90] (s.south);
\draw[->,thick] (fork) to[out=135,in=-45] (s.south east);
\node at (2.7,-.7) {$5$};
\end{scope}
\end{tikzpicture}
\caption{An instance $I$ of \PAFP with a safe path $s,a,t$ (left) and the corresponding PAC in $N_I$ (right).}
\label{fig:example}
\end{figure}
\end{proof}

We now turn to the converse direction:
\begin{lemma}\label{lem:backward}
If $N_I$ contains a PAC, then $G$ has a safe $s$--$t$
path.
\end{lemma}
\begin{proof}
Let $(X,Q)$ be a PAC in $N_I$, witnessed by a flow vector $\vec v$.

We claim that $X$ cannot contain both species $u$ and $\bar u$, for any forbidden pair $\{u,\bar u\}\in\mathcal P$. Recall that in our formulation of \PAFP, the forbidden pairs do not involve the source and target vertices $s$ and $t$. 
If both $u$ and $\bar u$ are in $X$, then all reactions from $Q$ keep constant the total amount of $u+\bar u$, so the net production of $u+\bar u$ is zero. This contradicts the existence of $\vec v$ that would allow for a  strict production of all species from $X$. Hence $X$ contains at most one species from each forbidden pair.

This means that any reaction from $Q$, when restricted to the selected species $X$, is of the form $A\rev B$, except for the amplifier $t\rev 2s$.

It is known from \cite{blokhuis2020universal} that in a PAC, after orientation according to the witness\footnote{It is shown in \cite{kosc2025thermodynamic} that the orientation is unique.}, every species must be a reactant exactly once, and the subnetwork is connected. It is clear that the amplifier $t\rev 2s$ is the only reaction that can allow for self-amplification, otherwise the total amount of species would be preserved by all reactions. Thus, the only possible shape for the PAC $(X,Q)$ is a simple path of incidence reactions, closed by the amplifier $t\to 2s$. This is exactly the shape of the PAC constructed in Lemma~\ref{lem:forward} from a safe $s$--$t$ path.
We obtain that the selected species $X$ form a safe $s$--$t$ path in $G$, as desired.
\end{proof}

This completes the proof of Lemma~\ref{lem:reduction}, and therefore of Theorem~\ref{thm:npcomplete}.

\section{Conclusion}
We have shown that the problem of detecting an autocatalytic core in a reaction network is \NP-complete, even when all reactions are reversible. The proof is a reduction from the undirected path problem with disjoint forbidden pairs, and the PACs used in the reduction are simple loops with a forked reaction, corresponding to type-I autocatalytic cores in the classification of \cite{blokhuis2020universal}.
This strengthens the results of Kosc et al.~\cite{kosc2025thermodynamic}, who proved \NP-completeness for a more constrained version of the problem.
It further justifies in particular the use of tools such as SAT- or SMT-solvers to list all PACs in a given network, as reported in \cite{kosc2025thermodynamic} for the software EmergeNS \cite{emergens}.

\bibliographystyle{unsrtnat}
\bibliography{pac_references}
\end{document}